\documentclass[12pt]{article}
\usepackage{amsmath}
\usepackage{amssymb}
\usepackage{epsfig}
\usepackage{graphicx}
\begin{document}

\newcommand{\be}{\begin{equation}}
\newcommand{\ba}{\begin{eqnarray} \nonumber }
\newcommand{\ee}{\end{equation}}
\newcommand{\ea}{\end{eqnarray}}

\title{{\bf THE DYNAMICS OF FINANCIAL MARKETS -- 
MANDELBROT'S MULTIFRACTAL CASCADES, AND BEYOND}}

\author{Lisa Borland$^1$, Jean-Philippe Bouchaud$^{2,3}$ \\
Jean-Fran\c{c}ois Muzy$^4$, Gilles Zumbach$^5$}
\maketitle
\small{$1$ Evnine-Vaughan Associates, Inc., 456 Montgomery Street, 
Suite 800, San Francisco, CA 94104,  USA\\
$2$ Science \& Finance, Capital Fund Management, 6 Bd Haussmann, 75009 Paris, France\\
$3$ Service de Physique de l'\'Etat Condens\'e, Centre d'\'etudes de Saclay,  Orme des Merisiers, 
91191 Gif-sur-Yvette Cedex, France\\
$4$ Laboratoire SPE, CNRS UMR 6134, Universit\'e de Corse, 20250 Corte, France
\\
$5$ Consulting in Financial Research, Chemin Charles Baudouin 8, 1228 Saconnex d'Arve, Switzerland}

\begin{abstract}
This is a short review in honor of B. Mandelbrot's 80st birthday, to appear in Wilmott magazine.
We discuss how multiplicative cascades and related multifractal ideas might be relevant to model the main statistical 
features of financial time series, in particular the intermittent, long-memory nature of the volatility. We describe
in details the Bacry-Muzy-Delour multifractal random walk. We point out some inadequacies of the
current models, in particular concerning time reversal symmetry, and propose 
an alternative family of multi-timescale models, 
intermediate between GARCH models and multifractal models, that seem 
quite promising.
\end{abstract}

\newpage

\section{Introduction}

Financial time series represent an extremely rich and fascinating source of questions, for they store a 
quantitative trace of human activity, sometimes over hundreds of years. 
These time series, perhaps surprisingly, turn out to reveal a very rich and particularly non trivial statistical 
texture, like fat tails and intermittent volatility bursts, have been by now well-characterized by an uncountable 
number of empirical studies, initiated by Benoit Mandelbrot more than fourty years ago \cite{MandelB}. Quite interestingly, 
these statistical anomalies are to some degree {\it universal}, i.e, common across different assets 
(stocks, currencies, commodities, etc.), regions (U.S., Europe, Asia) and epochs (from wheat prices in the 18th 
century to oil prices in 2004). The statistics of price changes is very far from the Bachelier-Black-Scholes random 
walk model which, nevertheless, is still today the central pillar of mathematical finance: the vast majority of books on 
option pricing written in the last ten years mostly focus on Brownian motion models and up until recently 
very few venture 
into the wonderful world of non Gaussian random walks \cite{Montroll}. This is the world, full of bushes, gems and 
monsters, that Mandelbrot started exploring for us in the sixties, charting out its scrubby paths, on which droves 
of scientists --
admittedly with some delay -- now happily look for fractal butterflies, heavy-tailed marsupials or long-memory 
elephants. One can only be superlative about his relentless efforts to look at the world with new goggles, and to
offer the tools that he forged to so many different communities: mathematicians, physicists, geologists, 
meteorologists, fractologists, economists -- and, for our purpose, financial engineers. In our view, one of Mandelbrot's 
most important methodological messages is that one should {\it look} at data, charts and graphs in order 
to build one's intuition, 
rather than trust blindly the result of statistical tests, often inadequate and misleading, in particular in the 
presence of non Gaussian effects \cite{Mandel-quote}. This visual protocol is particularly relevant when modeling 
financial time series: as discussed below, well chosen graphs often allow one to identify important 
effects, rule out an idea or construct a model. This might appear as a trivial statement but, as testified by Mandelbrot 
himself, is not: he fought all his life against the Bourbaki principle that pictures and graphs betray \cite{Mandel-quote}. 
Unreadable 
tables of numbers, flooding econometrics papers, perfectly illustrate that figures remain suspicious in many 
quarters. 

The aim of this paper is to give a short review of the stylized facts of financial time series, and of how 
multifractal stochastic volatility models, inherited from Kolmogorov's and Mandelbrot's work in the context of 
turbulence, fare 
quite well at reproducing many important statistical features of price changes. We then discuss some inadequacies of the
current models, in particular concerning time reversal symmetry, and propose an alternative 
family of multi-timescale models, intermediate
between GARCH models and multifractal models, that seem quite promising. We end on a few words about how these models 
can be used in practice for volatility filtering and option pricing. 

\section{Universal features of price changes}

The modeling of random fluctuations of asset prices is of obvious
importance in finance, with many applications to risk control, derivative pricing and systematic trading. 
During the last decade, the availability of high frequency 
time series has promoted intensive statistical studies
that lead to invalidate the classic and popular ``Brownian walk'' model, as anticipated by Mandelbrot 
\cite{cotton}.
In this section, we briefly review the main
statistical properties of asset prices that
can be considered as universal, in the sense that they are
common across many markets and epochs \cite{Guillaume,MS,Book}.

If one denotes $p(t)$ the price of an asset at time $t$, the return $r_{\tau}(t)$, at time $t$ and 
scale $\tau$ is simply the
relative variation of the price from $t$ to $t+\tau$:
$r_\tau(t) = \left[{p(t+\tau)-p(t)}\right]/{p(t)}\simeq \ln p(t+\tau) -\ln p(t)$.

The simplest ``universal'' feature of financial time series, uncovered by Bachelier in 1900,
is the linear growth of the variance of the return 
fluctuations with time scale. More precisely, if $m \tau$ is the mean return at scale $\tau$, 
the following property holds to a good approximation:
\begin{equation}
\langle \left(r_\tau(t)-m\tau \right)^2 \rangle_e \simeq \sigma^2 \tau,
\end{equation}
where $\langle ... \rangle_e$ denotes the empirical average. 
This behaviour typically 
holds for $\tau$ between a few minutes to a few years, 
and is equivalent to the
statement that  
relative price changes are, to a good approximation, 
{\it uncorrelated}. 
Very long time scales (beyond 
a few years) are difficult to investigate, in particular because the average
drift $m$ becomes itself time dependent, but there are systematic effects suggesting 
some degree of mean-reversion on these long time scales.

\begin{figure}[t]
\vskip 0.3cm
\epsfig{file=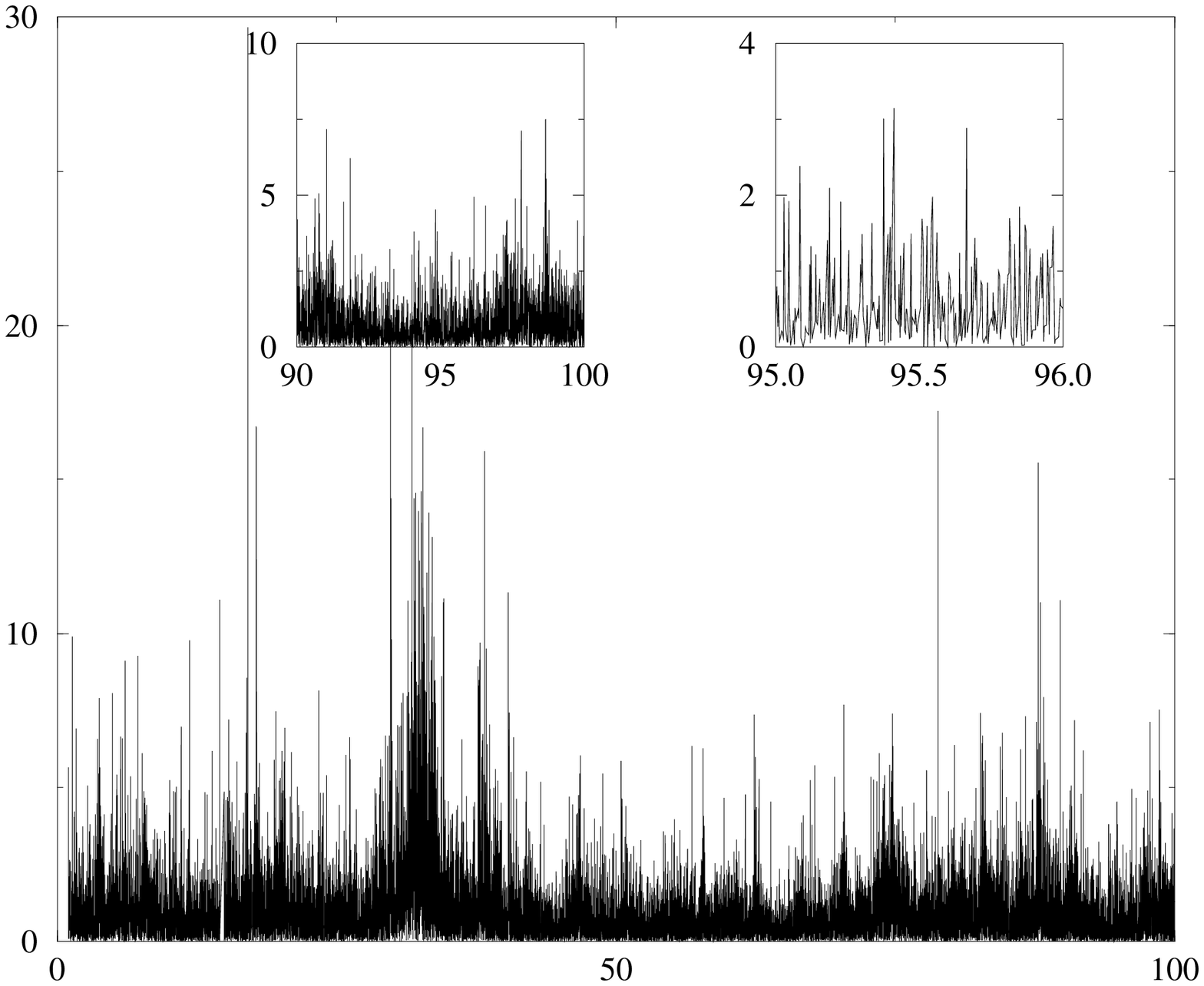,width=5cm,angle=270}\hspace{1cm}
\epsfig{file=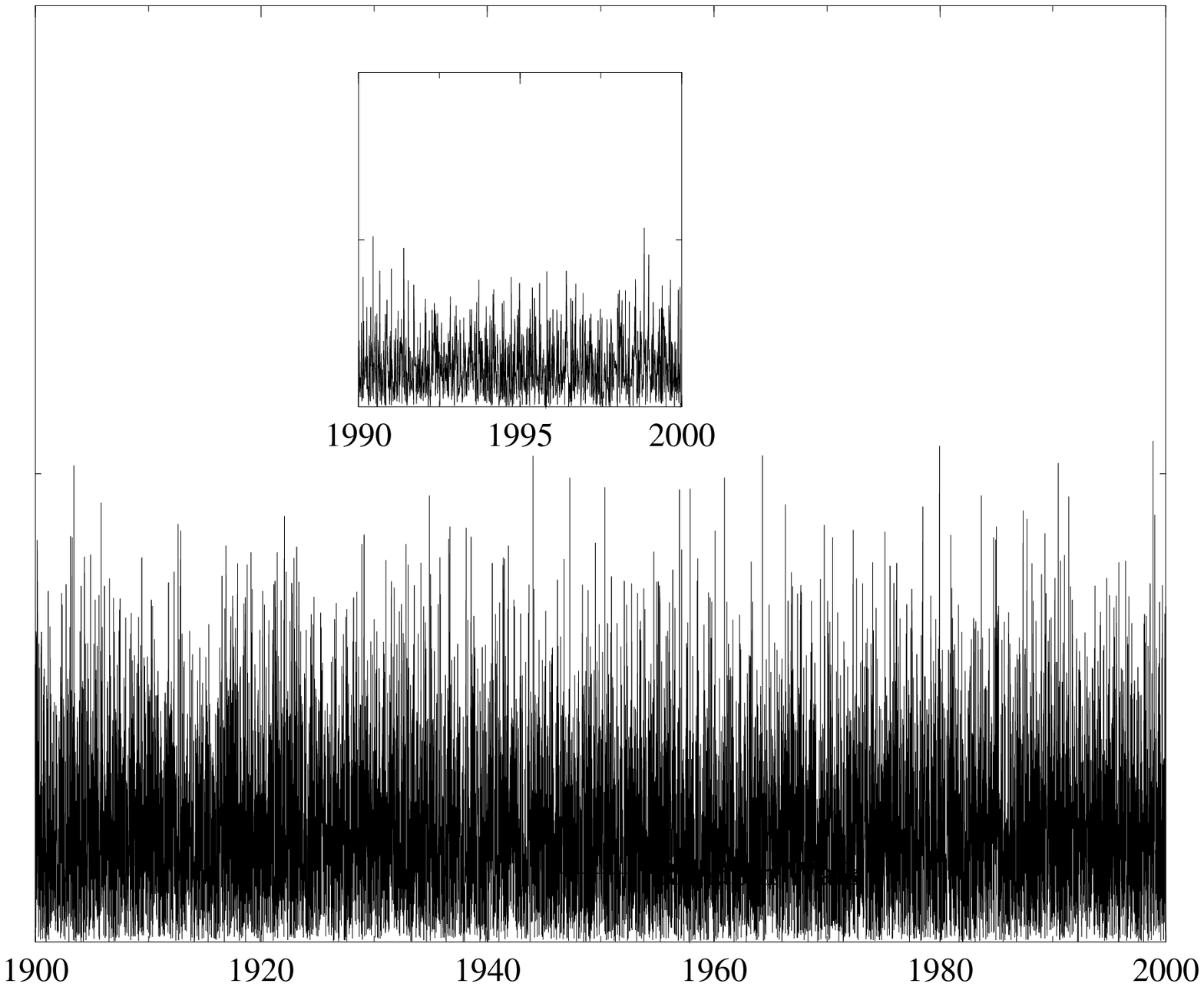,width=5cm,angle=270}
\caption{\small  
1-a: Absolute value of the daily price returns for the Dow-Jones index over a century (1900-2000), and zoom on different
scales (1990-2000 and 1995). Note that the volatility can remain high for a few years (like in the early 1930's) or for 
a few days. This volatility clustering can also be observed on high frequency (intra-day) data. 1-b: Same plot for 
a Brownian random walk, which shows a featureless pattern in this case. \label{fig2} }
\end{figure}

\begin{figure}[t]
\begin{center}
\epsfig{file=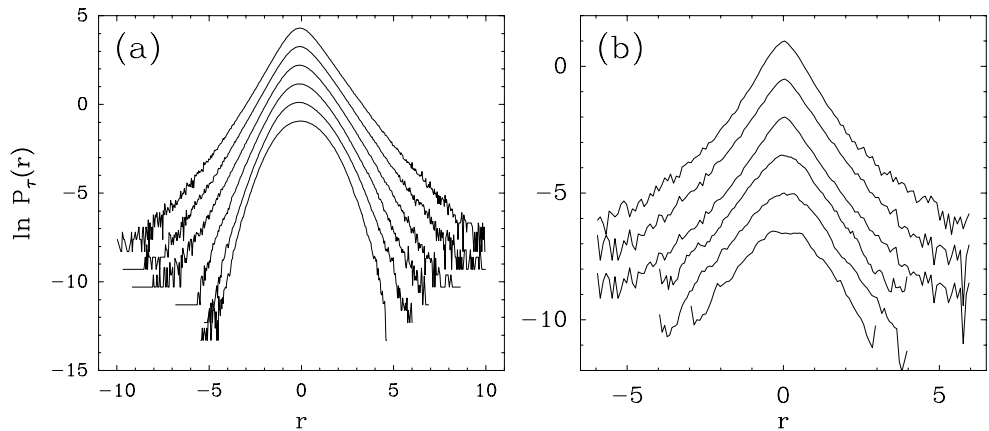,width=13cm}
\caption{\small Continuous deformation
of turbulent velocity increments and financial returns
distributions from small (top) to large (bottom) scales.  
2-a: Standardized probability distribution functions
of spatial velocity increments at different length scales 
in a high Reynolds number wind tunnel turbulence experiment. 
The distributions are plotted in logarithmic scale so that
a parabola corresponds to a Gaussian distribution.
2-b: Standardized p.d.f. of S\&P 500 index returns
at different time scales from few minutes to one month.
Notice that because of sample size limitation, 
the noise amplitude is larger
than in turbulence. \label{fig3} }
\end{center}
\end{figure}

The volatility $\sigma$ is the simplest quantity that measures the amplitude
of return fluctuations and therefore that quantifies the risk
associated with some given asset. Linear growth of the variance 
of the fluctuations with time is typical of the Brownian motion, 
and, as mentioned above, was proposed as a model of market fluctuations by Bachelier.\footnote{In 
Bachelier's model, absolute price changes, rather than relative returns, were considered; there 
are however only minor differences between the two at short time scales, $\tau <$ 1 month, but see the
detailed discussion \cite{Book}, Ch. 7.}
In this model, returns are not only uncorrelated, 
as mentioned above, but actually {\it independent} and identical
Gaussian random variables. However, this model 
completely fails to capture 
other statistical features of financial markets that even a rough
analysis of empirical data allows one to identify, at least qualitatively: 

\begin{itemize}

\item[(i)]
The distribution of returns is in fact strongly non Gaussian
and its shape continuously depends on the return period $\tau$:
for $\tau$ large enough (around few months), one observes 
quasi-Gaussian distributions while for small $\tau$ values,
the return distributions have a strong kurtosis (see Fig. \ref{fig3}).
Several careful studies suggest that 
these distributions can be characterized by Pareto (power-law) 
tails $|r|^{-1-\mu}$ with an exponent $\mu$ in the range $3 - 5$ even for liquid markets
such as the US stock index, major currencies, or interest rates 
\cite{Gopi1,Lux,Guillaume,Longin}. Note that $\mu > 2$, and excludes an early suggestion by 
Mandelbrot that security prices could be described by L\'evy stable laws. 
Emerging markets, however, have even more extreme tails, with an exponent $\mu$ that can be less than $2$ -- 
in which case the volatility is formally infinite -- as found by Mandelbrot in his famous study of 
cotton prices \cite{cotton}. 
A natural conjecture is that as markets become more liquid, the value of 
$\mu$ tends to increase (although illiquid, yet 
very regulated markets, could on the contrary show 
truncated tails because of artificial trading rules \cite{India}).   

\item[(ii)] 
Another striking feature is the intermittent and correlated 
nature of return amplitudes. At some given time period $\tau$, 
the volatility is a quantity that 
can be defined locally in various ways: the simplest ones
being the square return $r^2_\tau(t)$ or the absolute return
$|r_\tau(t)|$, or a local moving average of these quantities. The volatility signals
are characterized by self-similar outbursts (see Fig. \ref{fig2})
that are similar to intermittent variations of dissipation
rate in fully developed turbulence \cite{Frisch}.
The occurrence of such bursts are strongly correlated and
high volatility periods tend to persist in time.
This feature is known as {\it volatility clustering} 
\cite{volfluct1,volfluct2,MS,Book}. 
This effect can analyzed more quantitatively:  
the temporal correlation function
of the (e.g. daily) volatility can be 
fitted by an inverse power of the lag, with a rather small exponent in
the range $0.1 - 0.3$ \cite{volfluct2,PCB,Cizeau1,Book,mrw1}.

\item[(iii)]
One observes a non-trivial ``multifractal'' scaling \cite{Ghasg,Mandel3,Multif1,Multif2,mrw1,Lux2,CF0}, 
in the sense that higher moments of price changes scale anomalously with time:
\begin{equation}\label{fund}
M_q(\tau)=\langle \left|r_\tau(t)-m\tau \right|^q \rangle_e \simeq A_q \tau^{\zeta_q},
\end{equation}
where the index $\zeta_q$ is not equal to the Brownian walk value $q/2$.
As will be discussed more precisely below, this behaviour is intimately 
related to the intermittent nature of the volatility process.

\item[(iv)]
Past price changes and future 
volatilities are negatively correlated, at least on stock markets. This is 
the so called {\em leverage effect}, 
which reflects the fact that markets become 
more active after a price drop, and tend to calm down when the price rises. This
correlation is most visible on stock indices \cite{Matacz}. This leverage effect 
leads to an anomalous negative skew 
in the distribution of price changes as a function of time. 

\end{itemize}

The most important message of 
these empirical studies is that price changes behave very
differently from the simple geometric Brownian motion: 
extreme events are much more probable, and interesting non
linear correlations (volatility-volatility and price-volatility) are observed. 
These ``statistical anomalies'' are very
important for a reliable estimation of 
financial risk and for quantitative option pricing and hedging, for which one 
often requires an accurate model that captures
return statistical features on different time horizons $\tau$. 
It is rather amazing to remark that empirical properties
(i-iv) are, to some extent, also observed
on experimental velocity data in fully developed turbulent flows (see Fig. 
\ref{fig3}).
The framework of scaling theory and multifractal analysis,
initially proposed to characterize turbulent signals by Mandelbrot and others \cite{Frisch},
may therefore be well-suited to further characterize
statistical properties of price changes on different time periods. 

\section{From multifractal scaling to cascade processes} 

\subsection{Multi-scaling of asset returns}

\begin{figure}[t]
\begin{center}
\epsfig{file=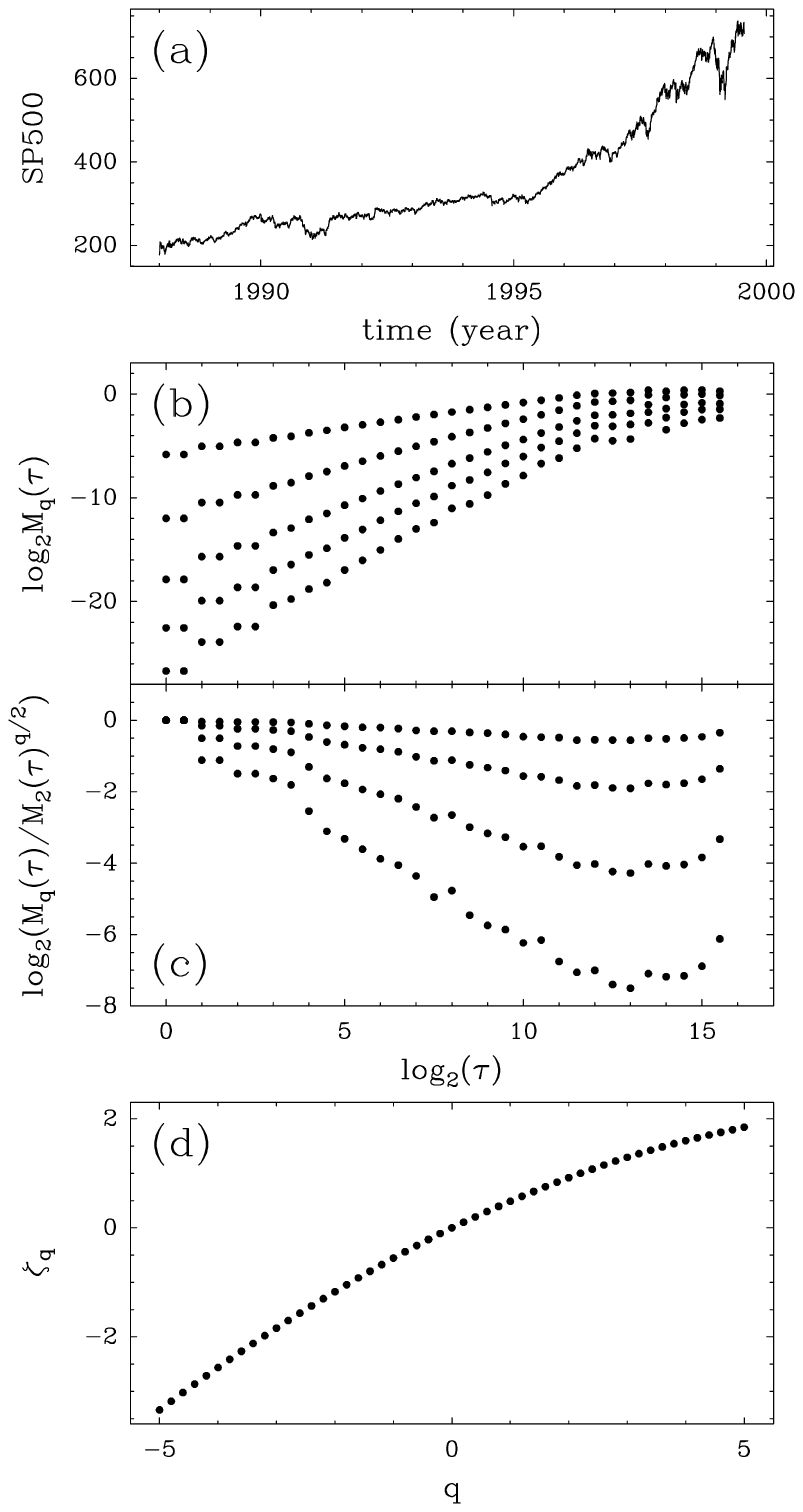,width=7cm}\hspace{1cm}
\caption{\small Multifractal scaling analysis of S\&P 500 returns.  
3-a: S\&P 500 index price series sampled at a 5 minutes rate.
3-b: First five absolute moments of the index (defined
in Eq. (\ref{fund})) as a function of the time period $\tau$ 
in double logarithmic representation. For each moment,
a linear fit of the small scale behavior provides an estimate of $\zeta_q$.
3-c: Moment ratios $M_q(\tau)/M_2(\tau)^{q/2}$ in log-log representation.
Such curves would be flat for a geometric Brownian process. 
3-d: $\zeta_q$ spectrum estimate versus $q$. Negative $q$ values
are obtained using a wavelet method as defined in e.g. \cite{mmto}.
This function is well fitted by a Kolmogorov log-normal spectrum (see text). 
\label{fig4} }
\end{center}
\end{figure}

For the geometric Brownian motion, or for the L\'evy stable processes first suggested 
by Mandelbrot as an alternative, the return distribution 
is identical (up to a rescaling) for all $\tau$. As emphasized in previous section,
the distribution of real returns is {\it not} scale invariant, but rather exhibits multi-scaling.
Fig. \ref{fig4}  illustrates the empirical multifractal analysis of
S\&P 500 index return. As one can see in Fig. \ref{fig4}(b), the 
scaling behavior (\ref{fund}) that corresponds to a linear dependence
in a log-log representation of the absolute moments versus the time
scale $\tau$, is well verified over some range of time scales (typically 3 
decades).

The multifractal nature of the index fluctuations
can directly be checked in Fig. \ref{fig4}(c), where 
one sees that the moment ratios strongly
depend on the scale $\tau$.
The estimated $\zeta_q$ spectrum (Fig. \ref{fig4}(d)) has a concave 
shape that is well fitted by the parabola: 
$\zeta_q = q(1/2+\lambda^2)-\lambda^2 q^2/2$ with $\lambda^2 \simeq 0.03$.
The coefficient $\lambda^2$ that quantifies the curvature
of the $\zeta_q$ function is called, in the framework
of turbulence theory, the {\em intermittency coefficient}.
The most natural way to account for the 
multi-scaling property (\ref{fund}) is through the notion
of cascade from coarse to fine scales.

\subsection{The cascade picture}
As noted previously, for the geometric Brownian motion,
the return probability distributions at different time periods $\tau$
are Gaussian and thus differ only by their width that is  
proportional to $\sqrt{\tau}$. If $x(t)$ is a 
Brownian motion, this property can be shortly written as
\begin{eqnarray}
  r_{\tau}(t) & \operatornamewithlimits{=}_{law} & \sigma_{\tau} \varepsilon(t) \\
   \sigma_{s \tau}  & = & s^{1/2} \sigma_{\tau}
\end{eqnarray}
where $\operatornamewithlimits{=}_{law}$ means that the two
quantities have the same probability distribution,  
$\varepsilon(t)$ is a standardized Gaussian white noise
and $\sigma_{\tau}$ is the volatility at scale $\tau$.
When going from some scale $\tau$ to the scale $s \tau$,
the return volatility is simply multiplied by $s^{1/2}$.
The cascade model assumes such a multiplicative rule but 
the multiplicative factor is now a random variable
and the volatility itself becomes a random process $\sigma_{\tau}(t)$:
\begin{eqnarray}
\label{rsh}
  r_{\tau}(t) & \operatornamewithlimits{=}_{law} & \sigma_{\tau}(t)
  \varepsilon(t) \\
\label{casc}
   \sigma_{s \tau} (t) & \operatornamewithlimits{=}_{law} &
W_{s} \sigma_{\tau}(t)
\end{eqnarray}
where the law of $W_{s}$ depends only on the scale
ratio $s$ and is independent of $\sigma_{\tau}(t)$. 
Let $T$ be some coarse time period and let $s < 1$. 
Then, by setting $W = e^{\xi}$,
and by iterating equation (\ref{casc}) $n$ times, one obtains:
\begin{equation}
  \sigma_{s^{n} T}(t)   \operatornamewithlimits{=}_{law} W_{s^{n}}
   \sigma_{T}(t) \operatornamewithlimits{=}_{law} 
   e^{\sum_{\ell=1}^n \xi_{\ell, s}} \sigma_{T}(t)
\end{equation}
Therefore, the {\em logarithm} of the volatility at some
fixed scale $\tau = s^n T$, can be written as
a sum of an arbitrarily large number $n$ of independent, identically
distributed random 
variables. Mathematically, this means that the logarithm 
of the volatility (and hence $\xi$, the logarithm of the ``weights'' $W$)
belongs the the class of the so-called
infinitely divisible distributions \cite{feller}. 
The simplest of such distributions (often invoked
using the central limit theorem) is the Gaussian law.
In that case, the volatility is a log-normal random
variable. As explained in the next subsection, this
is precisely the model introduced by Kolmogorov
in 1962 to account for the intermittency of
turbulence \cite{k62}. It can be proven that the random 
cascade equations (\ref{rsh}) and (\ref{casc}) directly
lead to the deformation of return probability distribution
functions as observed in Fig. \ref{fig3}.
Using the fact that $\xi_s$ is a Gaussian variable of mean $\mu \ln(s)$ 
and variance $\lambda^2 \ln(s)$, one can compute the absolute moment of order $q$ of
the returns at scale $\tau = s^{n} T$ ($s< 1$). One finds:
\begin{equation}
 \langle | r_{\tau}(t) |^q \rangle   =  
 A_q \left( \frac{\tau}{T} \right)^{\mu q -\lambda^2 q^2/2}
\end{equation}
where $A_q= \langle | r_{T}(t) |^q \rangle$. 

Using a simple multiplicative cascade, we thus have recovered
the empirical findings of previous sections.
For log-normal random weights $W_s$, the return process
is multifractal with a parabolic $\zeta_q$ scaling spectrum:  
$\zeta_q = q \mu -\lambda^2 q^2/2$ where the parameter $\mu$ is related to
the mean of $\ln W_s$ and the curvature $\lambda^2$ (the
intermittency coefficient) is related to the variance
of $\ln W_s$ and therefore to the variance of the log-volatility $\ln(\sigma_{\tau})$:
\begin{equation}
\label{var}
 \langle(\ln \sigma_\tau(t))^2 \rangle - \langle\ln \sigma_\tau(t) \rangle^2 
= -\lambda^2 \ln(\tau) + V_0
\end{equation}
The random character of $\ln W_s$ is therefore directly 
related to the intermittency of the returns.

\section{The multifractal random walk}

\begin{figure}[t]
\begin{center}
\epsfig{file=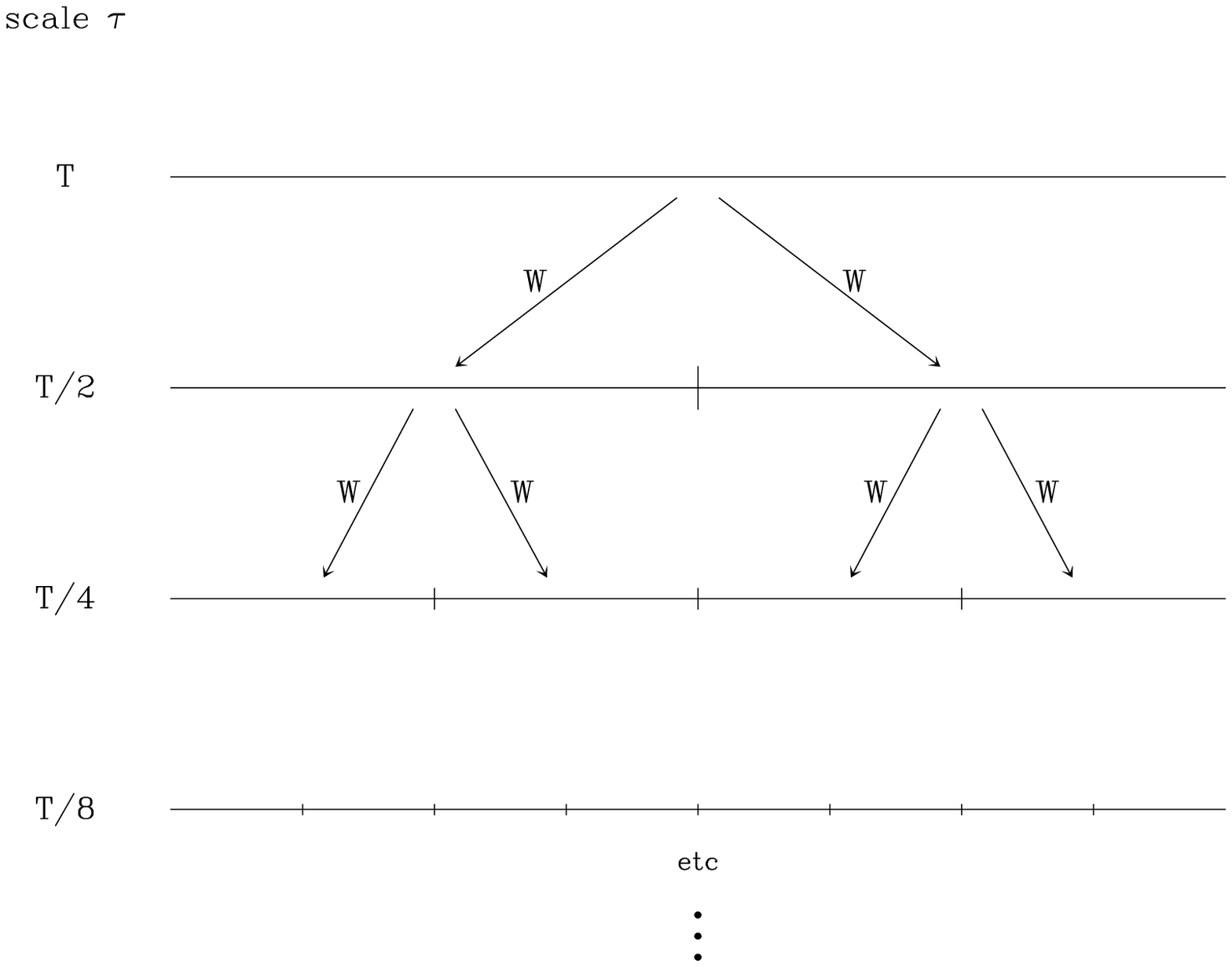,width=10cm}
\caption{\small Multiplicative construction of
a Mandelbrot cascade. One starts at the coarsest scale $T$ and
constructs the volatility fluctuations at fine scales
using a recursive multiplication rule. The variables $W$
are independent copies of the same random variable. \label{fig5} }
\end{center}
\end{figure}

The above cascade picture assumes that the volatility 
can be constructed as a {\em product} of random variables associated with
different time scales. In the previous section, we exclusively focused on 
return probability distribution functions (mono-variate laws)
and scaling laws associated with such models.
Explicit constructions of stochastic processes which marginals
satisfy a random multiplicative rule, were first introduced
by Mandelbrot \cite{mandel} and are known as Mandelbrot's
cascades or random multiplicative cascades. 
The construction of a Mandelbrot cascade, 
illustrated in Fig. \ref{fig5}, 
always involves a discrete scale ratio $s$ (generally $s=1/2$).
One begins with the volatility at the coarsest scale and proceeds in
a recursive manner to finer resolutions:
The $n$-th step of the volatility construction 
corresponds to scale $2^{-n}$ and 
is obtained from the $(n-1)$-th step by multiplication
with a positive random process $W$ the law of which does not depend on $n$. More
precisely, the volatility in each
subinterval is the volatility of its parent interval
multiplied by an independent copy of $W$.

Mandelbrot cascades are considered to be the paradigm of 
multifractal processes and have been extensively 
used for modeling scale-invariance properties in many fields, in 
particular statistical finance by Mandelbrot himself \cite{Mandel3,Lux2}.
However, this class of 
models presents several drawbacks: (i) they involve a preferred
scale ratio $s$, (ii) they are not stationary and (iii) they violate causality. 
In that respect, it is difficult to see how such models could arise from a
realistic (agent based) description of financial markets.

Recently, Bacry, Muzy and Delour (BMD) \cite{mrw1,mrw2} 
introduced a  model that does not possess any of the above limitations and
captures the essence of cascades through their correlation structure (see also
\cite{CF,Luxnew} for alternative multifractal models).

In the BMD model, the key ingredient is the volatility
correlation shape that mimics cascade features.
Indeed, as remarked in ref. \cite{Arn}, the tree like
structure underlying a Mandelbrot cascade, implies
that the volatility logarithm covariance decreases
very slowly, as a logarithm function, i.e.,
\begin{equation}
\label{covcasc}
\left \langle \ln(\sigma_{\tau}(t)) \ln(\sigma_{\tau}(t+\Delta t)) \right\rangle
-\left \langle \ln(\sigma_{\tau}(t)) \right\rangle^2 =  C_0-\lambda^2 \ln(\Delta t+\tau)
\end{equation}
This equation can be seen as a generalization of Kolmogorov
equation (\ref{var}) that describes only the behavior of the 
log-volatility variance.
It is important to note that such a logarithmic behaviour of the covariance
has indeed been observed for empirically estimated 
log-volatilities in various stock market data \cite{Arn}. 
The BMD model involves Eq. (\ref{covcasc}) within the continuous
time limit of a discrete stochastic volatility model.
One first discretizes time in units of 
an elementary time step $\tau_0$ 
and sets $t \equiv i \tau_0$. The volatility $\sigma_i$
at ``time'' $i$ is a log-normal 
random variable such that $\sigma_i=\sigma_0 \exp \xi_i$, where
the Gaussian process $\xi_i$ 
has the same covariance as in Eq. (\ref{covcasc}):
\begin{equation}\label{BMDlog}
\langle{\xi_i}\rangle = - \lambda^2 \ln (\frac{T}{\tau_0}) \equiv \mu_0 ; \qquad 
\langle{\xi_i \xi_j}\rangle - \mu_0^2= \lambda^2 \ln(\frac{T}{\tau_0}) - 
\lambda^2 \ln (|i-j|+1),
\end{equation}
for $|i-j| \tau_0 \leq T$. Here $T$
is a large cut-off time scale beyond which the volatility correlation
vanishes. In the above 
equation, the brackets stands for the mathematical expectation. 
The choice of the mean value $\mu_0$ is such that
$\langle{\sigma^2}\rangle = \sigma_0^2$. 
As before, 
the parameter $\lambda^2$ measures the intensity of volatility fluctuations
(or, in the finance parlance, the `vol of the vol'),
and corresponds to the intermittency parameter.
\begin{figure}[t]
\begin{center}
\epsfig{file=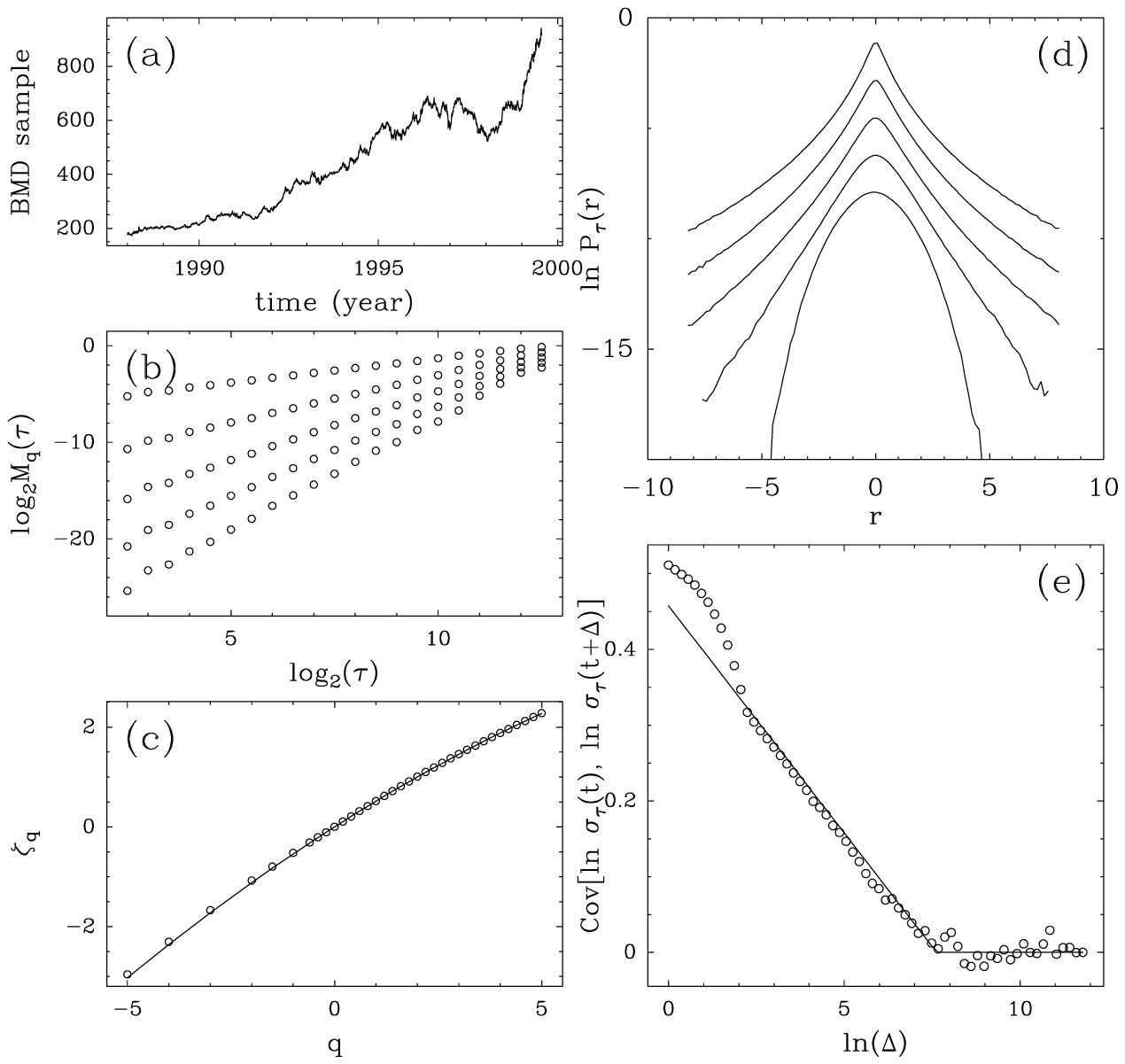,width=14cm}
\caption{\small Multifractal properties of BMD model. 5-a: Exponential of
a BMD process realization. The parameters have been adjusted in order
to mimic the features of S\&P 500 index reported in Fig. \ref{fig4}.
5-b: Multifractal scaling of BMD return moments for $q=1,2,3,4,5$.
5-c: Estimated $\zeta_q$ spectrum ($\circ$) compared with the
log-normal analytical expression (solid line).
5-d: Evolution of the return probability distributions across
scales, from nearly Gaussian at coarse scale (bottom) to
fat tailed law at small scales (top). (e) Log-volatility
covariance as a function of the logarithm of the lag $\tau$.
\label{fig6} }
\end{center}
\end{figure}

Now, the price returns are constructed as:
\begin{equation}
x\left[(i+1)\tau_0 \right]-x\left[i\tau_0\right] = r_{\tau_0}(i) 
\equiv \sigma_i \varepsilon_i = \sigma_0 e^{\xi_i} \varepsilon_i,
\end{equation}
where the $\varepsilon_i$ are a set of 
independent, identically distributed random variables of zero mean and 
variance equal to $\tau_0$. One also assumes that 
the $\varepsilon_i$ and the $\xi_i$ are independent (but see \cite{pb} where 
some correlations are introduced).
In the original BMD model, 
$\varepsilon_i$'s are Gaussian, and
the continuous time limit $\tau_0 = {\rm dt} \to 0$ is taken. 
Since $x = \ln p$, where $p$ is the price, the exponential of a 
sample path of the BMD model is plotted in Fig. \ref{fig6}(a), which can
be compared to the real price chart of \ref{fig3}(a).

The multifractal scaling properties of this 
model can be computed explicitly. Moreover, 
using the properties of 
multivariate Gaussian variables, one can get closed expressions
for all even moments $M_q(\tau)$ 
($q=2k$). 
In the case $q=2$ one trivially finds:
\begin{equation}
M_2(\tau = \ell \tau_0) = \sigma_0^2 \,  \ell \tau_0 \equiv \sigma_0^2 \tau,
\end{equation} 
independently of $\lambda^2$.  
For $q \neq 2$, one has to distinguish between the cases 
$q \lambda^2 < 1$ and $q \lambda^2 > 1$. 
For $q \lambda^2 < 1$, the corresponding moments 
are finite, and one finds, 
in the scaling region $\tau_0 \ll \tau \leq T$, a genuine multifractal
behaviour \cite{mrw1,mrw2}, for which Eq. ~(\ref{fund}) holds exactly. 
For $q \lambda^2 >  1$, on the other hand, the moments diverge, suggesting that the 
unconditional distribution of $x(t+\tau)-x(t)$ 
has power-law tails with an exponent $\mu = 1/\lambda^2$ (possibly multiplied 
by some slow function).
These multifractal scaling properties of BMD processes are numerically
checked in Figs. \ref{fig6}-a and \ref{fig6}-b where one
recovers the same features as in Fig. \ref{fig4} for the S\&P 500 
index. Since volatility correlations are absent for $\tau \gg T$, 
the scaling becomes that of a standard random walk, for which $\zeta_q=q/2$. 
The corresponding distribution of price returns thus becomes progressively 
Gaussian. An illustration of the progressive 
deformation of the distributions as $\tau$ increases,
in the BMD model is reported in Fig. \ref{fig6}-c. 
This figure can be directly compared to Fig. \ref{fig3}.
As shown in Fig. \ref{fig6}-e, this model also
reproduces the empirically observed logarithmic covariance
of log-volatility (Eq. (\ref{covcasc}). This functional
form, introduced at a ``microscopic'' level (at scale $\tau_0 \to 0$),
is therefore stable across finite time scales $\tau$.

To summarize, the BMD process is 
attractive for modeling financial time series since it reproduces most ``stylized facts'' reviewed in
section 2 and has exact multifractal properties as described in section 3. Moreover, this model
has stationary increments, does not exhibit any particular
scale ratio and can be formulated in a purely causal way: 
the log volatility $\xi_i$ can be expressed as a sum over ``past'' random
shocks, with a memory kernel that decays as the inverse square root of
the time lag \cite{MMS}. Let us mention that generalized stationary continuous cascades
with compound Poisson and infinitely divisible statistics have been recently contructed
and mathematically studied by Mandelbrot and Barral \cite{Barral}, and in \cite{mrwid}
However, there is no distinction between past and future in this model -- but see below.

\section{Multifractal models: empirical data under closer scrutiny} 

Multifractal models either postulate or predict a number of precise statistical
features that can be compared with empirical price series. We review here several
successes, but also some failures of this family of models, that suggest that the
century old quest for a faithful model of financial prices might not be over yet, despite 
Mandelbrot's remarkable efforts and insights. Our last section will be devoted to an idea
that may possibly bring us a little closer to the yet unknown ``final'' model. 

\subsection{Volatility distribution}

Mandelbrot's cascades construct the local volatility as a product of random volatilities 
over different scales. As mentioned above, this therefore leads to the local volatility
being log-normally distributed (or log infinitely divisible), an assumption that the BMD model 
postulates from scratch. 
Several direct studies of the distribution of the volatility are indeed compatible with log-normality;
option traders actually often think of volatility changes in {\it relative} terms, suggesting 
that the log-volatility is the natural variable. However, other distributions, such as an inverse gamma distribution 
fits the data equally well, or even better \cite{Mantegna,Book}. 

\subsection{Intermittency coefficient}

One of the predictions of the BMD model is the equality between the intermittency coefficient
estimated from the curvature of the $\zeta_q$ function and the slope of the log-volatility covariance 
logarithmic decrease. The direct study of various empirical log-volatility correlation functions, 
show that can they can indeed be fitted by a logarithm of the time lag, with a slope that is in the same
ball-park as the corresponding intermittency coefficient $\lambda^2$. The agreement is not perfect, though.
These empirical studies furthermore suggest that the integral time $T$ is on the scale of a few years \cite{mrw1}.

\subsection{Tail index}

As mentioned above, and as realized by Mandelbrot long ago through his famous ``star equation'' \cite{mandel}, 
multifractal models lead to power law tails 
for the distribution of returns. The BMD model predicts that the power-law index should be $\mu=1/\lambda^2$,
which, for the empirical values of $\lambda^2 \sim 0.03$, leads to a value of $\mu$ ten times larger than 
the empirical value of $\mu$, found in the range $3 - 5$ for many assets \cite{Gopi1,Lux,Guillaume}.
Of course, the random variable $\varepsilon$ could itself be non Gaussian and further fattens the tails. 
However, as recently realized by Bacry, Kozhemyak and Muzy \cite{Muzynew}, a kind of `ergodicity breaking' seems to take place in these
models, in such a way that the theoretical value $\mu=1/\lambda^2$ may not correspond to the most probable
value of the estimator of the tail of the return distribution for a given realization of the volatility, the
latter being much smaller than the former. So, even if this scenario is quite non trivial, multifractal models
with Gaussian residuals may still be consistent with fat tails. 

\subsection{Time reversal invariance}

Both Mandelbrot's cascade models and the BMD model are invariant under time reversal symmetry, meaning that
no statistical test of any nature can distinguish between a time series generated with such models and 
its time reversed. There are however various direct indications that real price changes are {\it not} time
reversal invariant. One well known fact is the ``leverage effect'' (point (iv) of Section 2, 
see also \cite{Matacz}), whereby a negative price change is on average followed by a volatility increase. 
This effect leads to a skewed distribution of returns, which in turn can be read as a skew in option
volatility smiles.  In the BMD model, one has by symmetry that all odd moments of the process vanish. 
A simple possibility, recently investigated in \cite{pb}, is to correlate negatively the variable $\xi_i$ 
with `past' values of the variables $\varepsilon_j$, $j < i$, through a kernel that decays as a power-law. 
In this case, the multifractality properties of the model are preserved, but the expression for $\zeta_q$ 
is different for $q$ even and for $q$ odd (see also \cite{kertecz}).

\begin{figure}[t]
\begin{center}

\epsfig{file=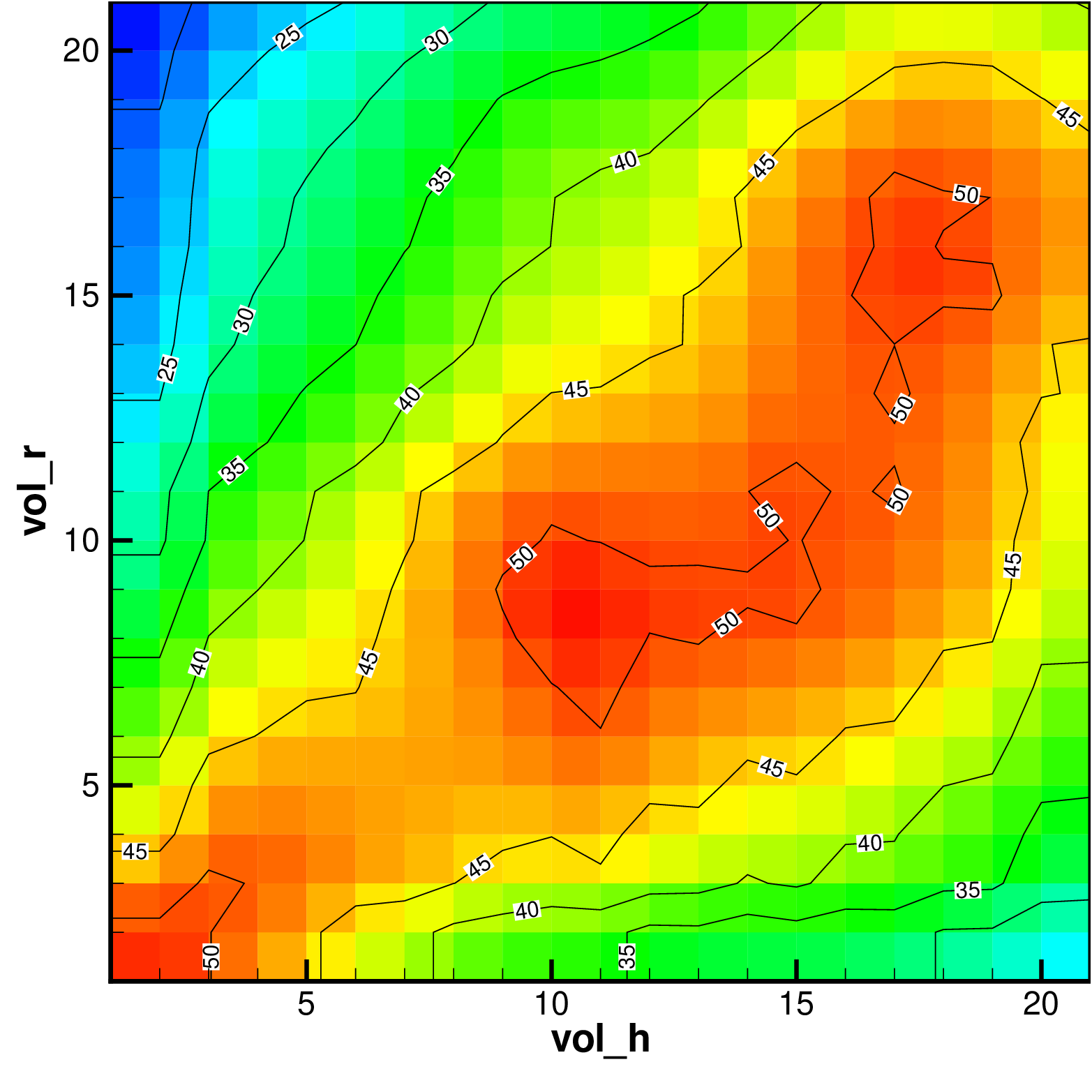,width=7cm}\hspace{1cm}
  \caption{\small The correlation between past historical volatilities $\sigma_h[\tau]$ 
  ($\ln(\tau)$ on the horizontal axis) and the future realized volatilities $\sigma_r[\tau']$ 
  ($\ln(\tau')$ on the vertical axis) for the USD/CHF foreign exchange time series.
  The units correspond to the time intervals: 1h 12min for (1), 4h 48min (4), 1d 14h 24min (10), 
  6d 9h 36min (14), and 25d 14h 24min (18).
  }
  \label{fig:volatilityCorrelation_USDCHF}
\end{center}
\end{figure}

\begin{figure}[t]
\begin{center}

\epsfig{file=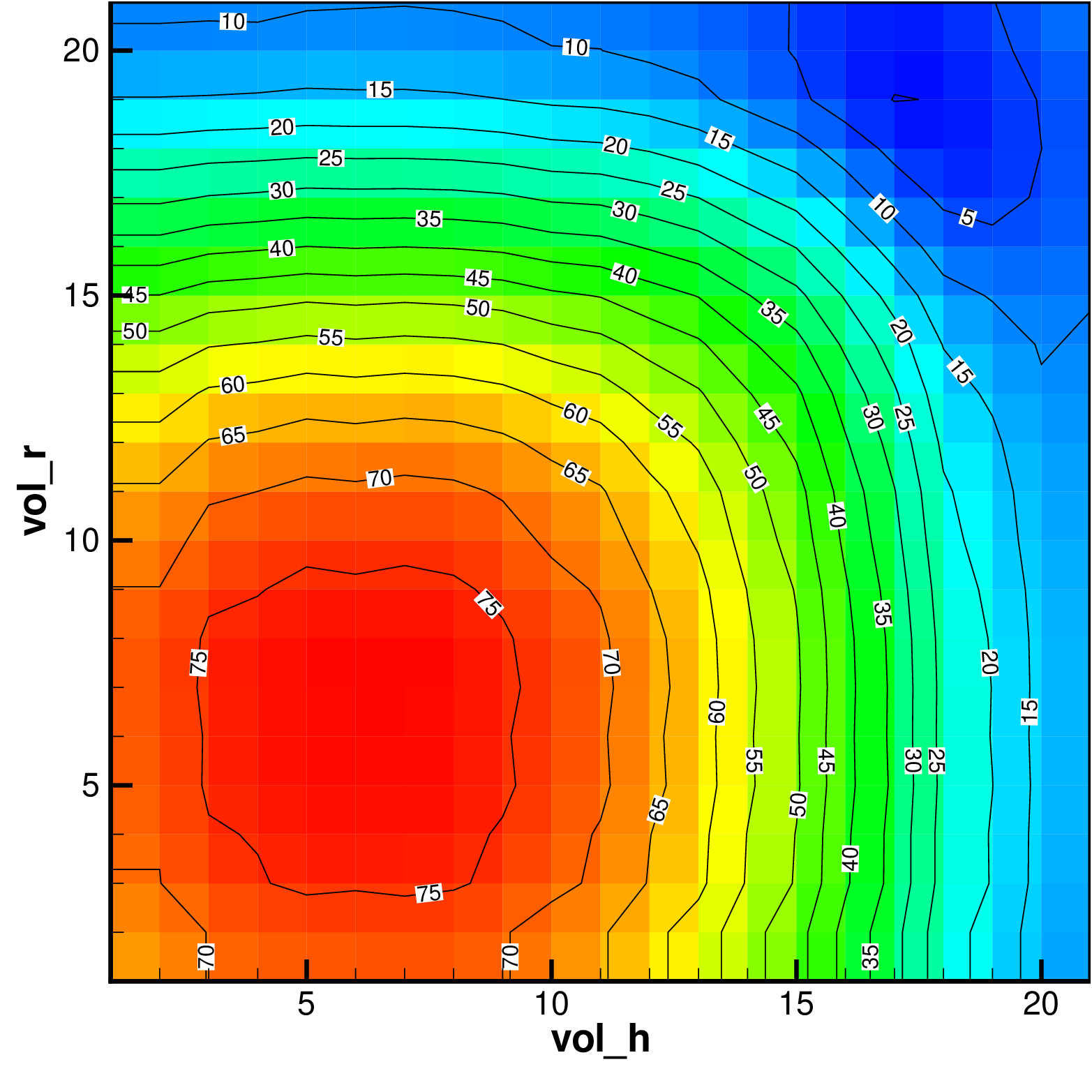,width=7cm}\hspace{1cm}
  \caption{\small As for Fig.~\ref{fig:volatilityCorrelation_USDCHF}, but for 
  a theoretical volatility cascade with Ornstein-Uhlenbeck partial log-volatilities $\xi_n$.
  The cascade includes components from 0.25 day to 4 days. Note the symmetry of this mug-shot, 
  to be contrasted with the previous figure. A similar, symmetric mug-shot would also obtain for 
  multifractal models..
  }
  \label{fig:volatilityCorrelation_expLM_SV}
\end{center}
\end{figure}

Another effect breaking the time reversal invariance
is the asymmetric structure of the correlations between the past and future volatilities
at different time horizons. This asymmetry is present in all time series, 
even when the above leverage effect is very small or inexistent,
such as for the exchange rate between two similar currencies
(e.g. Dollar vs. Euro).
The full structure of the volatility correlations is shown by the 
following ``mug shots'' introduced in \cite{Zumb}. 
These plots show how past volatilities measured on different time horizons
(horizontal axis) affect future volatilities, again on different time horizons 
(vertical axis). For empirical price changes, the correlations between 
past volatilities and future volatilities is asymmetric, 
as shown on Fig.~\ref{fig:volatilityCorrelation_USDCHF}.
This figure reveals clearly two very important features:
(a) the volatility at a given time horizon is affected 
mainly by the volatilities at longer time horizon 
(the ``cascade'' effect, partly discovered in \cite{Arn}), and
(b) correlations are strongest for time horizons 
corresponding to the natural human cycles, i.e: 
intra-day, one day, one week and one month.
The feature (a) points to a difference with the volatility cascade
used in turbulence, where eddies at a given scale break down
into eddies at the immediately smaller scale. 
For financial time series, the volatilities at
all time horizons feed the volatility at the shortest time horizon \cite{Zumb}.
The overall asymmetry around the diagonal of the figure measures
the asymmetry of the price time series under time reversal invariance.
For a time reversal symmetric process, 
these mug shots are symmetric.
An example of a very symmetric process 
is given in Fig.~\ref{fig:volatilityCorrelation_expLM_SV},
for a volatility cascade as given in Eq. (7).
The process for the ``log-volatility'' $\xi_n$ at the $n$-th scale obeys an 
Ornstein-Uhlenbeck process, 
with a characteristic time $s^{-n}$.
In order to obtain a realistic cascade of volatilities,
the (instantaneous) mean for the process $\xi_n$ is taken as the volatility 
at the larger scale $\xi_{n-1}$.
Yet, the resulting mug shot is very different from 
the empirical one, in particular it is exactly symmetric (at least theoretically).

Building a multifractal model that reproduces the observed time asymmetry 
is still an open problem.  There might however be other avenues, as we now discuss.

\section{Multi-timescale GARCH processes and statistical feedback}

The above description of financial data using multifractal, 
cascade-like ideas is still mostly phenomenological. 
An important theoretical issue is to understand the underlying 
{\it mechanisms} that may give rise to the remarkable statistical structure of the volatility
emphasized by multifractal models: (nearly) log-normal, with logarithmically decaying correlations.
Furthermore, from a fundamental point of view, 
the existence of {\it two} independent statistical processes, one 
for returns and another for the volatility, may not be very natural. 
In stochastic volatility models, such as the multifractal model,  the volatility has its 
own dynamics and sources of randomness, without any feedback from the price behaviour. 
The time asymmetry revealed by the above mug-shots however unambiguously shows that a strong feedback, 
beyond the leverage effect, exists in financial markets.

This time asymmetry, on the other hand, is naturally present in GARCH models, where the volatility is a (deterministic) 
function of the past price changes. The underlying intuition is that when recent price changes are large, 
the activity of market agents increases, in turn creating possibly large price changes.
This creates a non linear feedback, where rare events trigger more rare events,
generating non trivial probability distributions and correlations.
Most GARCH models describe the feedback as quadratic in past 
returns, with the following general form. The (normalized) return at time $i$, computed over a time interval $k\tau_0$ 
is given by $\tilde r_{i, k} = (x_i - x_{i-k})/\sqrt{k \tau_0}$. The general form of the volatility is:
\be\label{GARCH}
	\sigma_i^2 = \sigma_0^2 + \sum_{j < i} \sum_{k > 0} L(i-j; k) \tilde r_{j, k}^2  
\ee
where $\sigma_0$ is a `bare' volatility that would prevail in the 
absence of feedback, and $L(i-j;k)$ a kernel that describes how square returns on different time scales $k$, computed
for different days $j$ in the past, affect the 
uncertainty of the market at time $i$. [Usually, these processes are rather formulated through a set of recursive equations, 
with a few parameters. By expanding these recursion equations, one obtains the $AR(\infty)$ form given above.]

The literature on GARCH-like processes is huge, with literaly dozen of variations that have been proposed (for 
a short perspective, see \cite{Zumb,Zumb2}). Some of these models are very successful at reproducing most 
of the empirical facts, including the ``mug shot'' for the volatility correlation. An example, that was first 
investigated in \cite{Zumb} and recently rediscovered in \cite{LisaNew} as a generalization of previous work \cite{LisaOld}, 
is to choose $L(i-j;k) \equiv \delta_{i,j} K(k)$. 
This choice means that the current volatility is only affected by observed price variations between different dates $i-k$
in the past and today $i$. The quadratic dependence of the volatility on past returns is precisely the one found in the 
framework of the 
statistical feedback process introduced in \cite{LisaOld}, where the volatility is proportional to a negative power of the 
probability of past price changes -- hence the increased volatility after rare events. The time series resulting from the
above choice of kernel are found to exhibit fat, power-law tails, volatility bursts and long range memory \cite{LisaNew}. 
Furthermore, the distribution of volatilities is found to be very close to log-normal \cite{LisaNew}. 
Choosing $K$ to be a power-law of time \cite{Zumb,LBJPB} allows one to reproduce the power-law decay of volatility correlations 
observed in real data \cite{volfluct2,Cizeau1,Book,mrw1}, the power-law response of the volatility to 
external shocks \cite{MMS,kertecz2}, an apparent multifractal scaling \cite{BPM} and, most importantly
for our purpose, the shape of the mug-shots shown above \cite{Zumb}. Many of these results can be obtained 
analytically \cite{LisaNew,LBJPB}. Note that one could choose $K(k)$ to spike for the 
day, week, month time scales unveiled by the mug-shots \cite{Zumb,Zumb2}. Yet, a detailed systematic comparison of 
empirical facts against the predictions of the models, GARCH-like or multifractal-like, is still lacking. 

The above model can be generalized further by postulating a Landau-like expansion of the volatility as a function of past returns as:
\be\label{lisa}
\sigma_i^2 = \sigma_0^2 +  \sum_{k >0} K(k) G[\tilde r_{i,k}] 
\ee
with $G(r)=g_1 r + g_2 r^2 + ...$ \cite{LBJPB}. The case $g_1=0$ corresponds to the model discussed 
above, whereas $g_1 < 0$ allows one to reproduce the leverage effect and a negative skewness. 
Other effects, such as trends, could also be included 
\cite{Zumb3}. The interest of this type of models, compared to 
multifractal stochastic volatility models, is that their fundamental justification, 
in terms of agent based strategy, is relatively direct and plausible. 
One can indeed argue that agents use thresholds for their interventions 
(typically stop losses, entry points or exit points) that depend on the actual path of 
the price over some investment horizon, which may differ widely between different investors -- 
hence a power-law like behaviour of the kernels $K(k)$. 
Note in passing that, quite interestingly, Eq. (\ref{GARCH}) is rather at odds
with the efficient market hypothesis, which asserts that the price past history should have  
no bearing whatsoever on the behaviour of investors. 
Establishing the validity of Eq. (\ref{GARCH}) could have important repercussions on that front, too. 

\section{Conclusion: why are these models useful anyway ?}

For many applications, such as risk control and option pricing, we need a model that allows one to predict, as
well as possible, the future volatility of an asset over a certain time horizon, or even better, to predict that
probability distribution of all possible price {\it paths} between now and a certain future date. A useful model
should therefore provide a well defined procedure to {\it filter} the series past price changes, and to compute
the weights of the different future paths. Multifractal models offer a well defined set of answers to these questions, 
and their predictive power have been shown to be quite good \cite{CF,Muzypredict,Luxnew}. Other models, more in the spirit
of GARCH or of Eq. (\ref{lisa}) have been shown to also fare rather well \cite{HARCH,Zumb2,LBJPB,Zumb3}. 
Filtering past information within both frameworks is actually quite similar; it would be 
interesting to compare their predictive power in more details, with special care for non-Gaussian effects. 
Once calibrated, these models can in principle be used for VaR estimates and option pricing. However, their mathematical 
complexity do not allow for explicit analytical solutions (except in some special cases, see e.g. \cite{LisaOld}) and one 
has to resort to numerical, Monte-Carlo methods. The difficulty for stochastic volatility models or GARCH models in general 
is that the option price must be 
computed conditional to the past history \cite{pb2}, which considerably complexifies Monte-Carlo methods, 
in particular for path dependent options, or when non-quadratic hedging objectives are considered. In other words, both the
option price and the optimal hedge are no longer simple functions of the current price, but {\it functionals} of the 
whole price history. Finding operational ways to generalize, e.g. the method of Longstaff and Schwartz \cite{LS}, or 
the optimally hedged Monte-Carlo method of \cite{PBS} to account for this history dependence, seems to us a major
challenge if one is to extract the best of these sophisticated volatility models. The toolkit that Mandelbrot 
gave us to describe power-law tails and long-range memory is still incomplete: is there an Ito lemma to deal 
elegantly with multifractal phenomena? Probably not -- but as Paul Cootner pointed out long ago in his review of the 
famous cotton
price paper \cite{cotton}, Mandelbrot promised us with blood, sweat, toil and tears. After all, a 
faithful model, albeit difficult to work with, 
is certainly better than Black and Scholes's pristine, but often misleading, simplicity \cite{Welcome}. 

\section*{Acknowledgments}  
We thank E. Bacry, J. Delour and B. Pochart for sharing their results with us, T. Lux and M. Potters for 
important discussions, and J. Evnine for his continuous support and useful comments. 
This work arose from discussions between the authors during the meeting: ``Volatility 
of Financial Markets'' organized by the Lorentz Center, Leiden in October 2004.

\end{document}